\documentclass[twoside,fleqn]{article}
\usepackage{espcrc2MOD}
\usepackage{epsfig}

\newcommand{\be}{\begin{equation}}
\newcommand{\ee}{\end{equation}}
\newcommand{\bear}{\begin{eqnarray}}
\newcommand{\ear}{\end{eqnarray}}
\newcommand{\bea}{\begin{eqnarray*}}
\newcommand{\ea}{\end{eqnarray*}}

\newcommand{\gsim}{\mathrel{\vcenter
    {\hbox{$>$}\nointerlineskip\hbox{$\sim$}}}}
\newcommand{\bm}[1]{\mbox{\boldmath$#1$}}  
\renewcommand{\vec}{\bm}

\newcommand{\klgl}{\:\hbox to -0.2pt{\lower2.5pt\hbox{$\sim$}\hss}
 {\raise3pt\hbox{$<$}}\:}

\begin{document}
\bibliographystyle{personale}
\sloppy
\begin{titlepage}
\title{{\Large\bf Elastic Cross sections for high energy 
    hadron-hadron scattering}}
\author{E. R. Berger\thanks{
Supported by the Deutsche Forschungsgemeinschaft under 
    grant no. GRK 216/1-98} 
\address{Institut f\"ur Theoretische Physik der Universit\"at Heidelberg,\\
  Philosophenweg 16 \& 19, D-69120 Heidelberg}}
\begin{abstract}

This report discusses some results on
differential cross sections for high energy and small
momentum transfer elastic hadron-hadron scattering 
in QCD, using a functional integral approach.
In particulary a matrix cumulant
expansion for the 
vacuum expectation
values of lightlike Wegner-Wilson loops, which
governs the hadronic amplitudes, is presented.
The cumulants are evaluated using the model
of the stochastic vacuum.

\end{abstract}
\end{titlepage}

\maketitle

\section{Introduction}



We will discuss  here some results of \cite{naber}
for
elastic scattering of hadrons at high centre of
mass energy  $\sqrt s$ ($\sqrt s
\gsim
20$ GeV) and low momentum transfer squared $t$(say
$|t|{\mathrel{\vcenter
    {\hbox{$<$}\nointerlineskip\hbox{$\sim$}}}}
    \rm O (1\,{\rm GeV}^2)$). 
Such reactions are governed by soft,
nonperturbative interactions.

We are specifically interested in calculating elastic
differential cross sections (d$\sigma/dt$)
in terms of the description of diffractive hadron-hadron
scattering of \cite{dfk,na96}.
Based on the work of Nachtmann \cite{na91} it was shown 
there that the hadronic amplitudes are calculated from 
correlation functions of lightlike Wegner-Wilson loops.

\begin{figure}[!t]
  \unitlength1.0cm
  \begin{center}
    \begin{picture}(15.,8.8)

      \put(-0.4,3.0){
        \epsfysize=5.5cm

        \epsffile{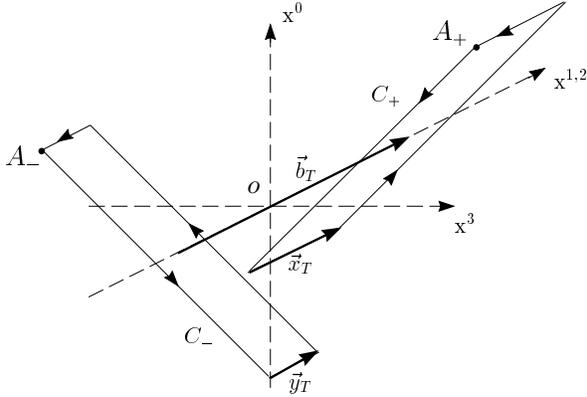} }

      \put(3.1,5.8){$o$}

    \end{picture}
  \end{center}
\vspace*{-4.2cm}
\caption[a]{ 
 The light-like Wegner-Wilson cut
loops in Minkowski space time,
$C_{\pm}$, consisting of two light like lines in the hyperplanes
$x_{\mp}=0$ and connecting pieces at infinity.
The loops are cut open at one corner, $A_-$ and $A_+$, respectively.}
  \label{loop}
\end{figure}

The result for meson-meson scattering where mesons are represented
as $q\bar{q}$-wave packets is
\begin{eqnarray}
  S_{fi} &=& \delta_{fi} + i(2\pi)^4 \delta (P_3+P_4-P_1-P_2) T_{fi},
  \nonumber\\
  T_{fi} &=& (-2is) \int d^2b_T {\exp}(i{\bm q}_T{\bm b}_T) \cdot
  \nonumber\\
  && \hspace{-1cm}
  \int d^2x_T \, d^2y_T \, w_{3,1}^M({\bm {x}}_T) \, w_{4,2}^M({\bm
    {y}}_T)
  \cdot
  \nonumber\\
  && \hspace{-1cm}
  \Big\langle W_+^M(\frac{1}{2} {\bm b}_T,{\vec x}_T)
  W_-^M(-\frac{1}{2} {\vec b}_T,{\vec y}_T)-1 \Big\rangle_G .
  \label{ampl}
\end{eqnarray}
%
%
Here the assumption is made that the $q$ and $\bar{q}$
share the longitudinal momentum of the meson
roughly in equal proportions.
The interpretation of (\ref{ampl}) and the symbols
occurring there is as follows.
The scattering amplitude is obtained by first considering the
scattering of quarks and antiquarks on a fixed gluon potential
and then summing over all
gluon potentials by path integration, indicated with the brackets
$\langle \; \rangle_G$. Travelling through a gluon potential the
quarks and antiquarks pick up non-abelian phase factors.
To ensure  gauge invariance the phase factors for
$q$ and $\bar q$ from the  same meson
are joined at large positive and negative
times, yielding lightlike Wegner-Wilson loops $W_{\pm}$
which are defined as
\bear
\label{wloop}
  &&\hspace{-0.69cm}
  W_{\pm}^M \equiv \frac{1}{3} \, {\rm{tr}} \, V(C_{\pm}),
  \nonumber\\
  &&\hspace{-0.69cm}
  V(C_{\pm}) =  {\rm P}\ {\rm exp} [ -ig\int_{\it{C_{\pm}}}
  dx^{\mu} \, G_{\mu}^{a}(x) \frac{\lambda^{a}}{2} ].
\ear
%
%
Here $V(C_{\pm})$ are non-abelian phase factors (connectors)
along cut loops $ C_{\pm}$ as sketched in Fig. \ref{loop}.
The transverse separation
between the centres of the loops is given by $\vec{b}_T$. The
vectors $\vec{x}_T$ and $\vec{y}_T$ give the extensions and
orientations of the loops in transverse space.
The path integration correlates
these loops and so causes the interaction.
The resulting loop-loop
correlation function  has to be integrated over all extensions and
orientations of the loops in transverse space  with a measure given by
the meson's overlap functions $w_{3,1}^M$ and
$w_{4,2}^M$ for which one has
to make a suitable ansatz.
To calculate (\ref{ampl}) a
Fourier transform over the impact parameter
$\vec{b}_T$ has to be done finally.
Meson-baryon and baryon-baryon scattering can be treated along 
similar lines \cite{dfk,na96}.

In what  follows baryons are treated as 
quark-diquark systems
where two quarks are close to each other on a scale
given by the proton radius. Then baryons do
act in a first approximation as colour dipoles in the same way as
mesons and we
can use the meson-meson amplitude (\ref{ampl}) also to describe 
meson-baryon and baryon-baryon scattering.
%
%
%
%
%
%

The colour dipole 
correlation function $\langle W_+^M W_-^M \rangle_G $ of (\ref{ampl})
will now be 
calculated, developing a matrix
cumulant expansion for products of 
Wegner-Wilson loops and using the model of the stochastic vacuum
(MSV) \cite{msv}, applied in Minkowski space after an analytic 
continuation from Euklidean space.

\section{Colour dipole correlation function}

We transform in a first step the line integrals $W_+^M$ and $W_-^M$
of $\langle W_+^M W_-^M \rangle_G $ into a surface integral
using the non-abelian Stokes theorem as presented in \cite{na96}.
Following \cite{dfk} we choose as surface the mantle of the double 
pyramid $P=P_+ + P_-$ which has $C_+ + C_-$ as boundary and $o$
as apex.
\begin{eqnarray}
  \label{stokes}
  &&\hspace{-0.69cm}
  \Big\langle W^M_+W^M_-\Big\rangle_G
    =\Big\langle\frac{1}{3}[{\rm tr} V(C_+)]\frac{1}{3}[{\rm tr}
  V(C_-]\Big\rangle_G\nonumber\\
  &&\hspace{1.5cm}
  = \Big\langle\frac{1}{3}[{\rm tr} V(P_+)]\frac{1}{3}[{\rm tr}
  V(P_-]\Big\rangle_G.
\ear
%
%
The matrices $V(P_{\pm})$ are surface ordered exponentials of field
strength tensors $\hat G$ parallel transported
to $o$. 

The main idea is now to interpret
the product of the two traces (tr) over $3\times 3$ matrices
in (\ref{stokes}) as one trace
$({\rm{Tr}}_2)$ acting in the $9$-dimensional
tensor product space carrying the product of two SU(3) quark
representations:
\begin{eqnarray}
  &&\hspace{-0.69cm}
  \Big\langle W_+^M \,   W_-^M \Big\rangle_G = 
  \frac{1}{9} \, {\rm{Tr}}_2 \Big\{
  \nonumber\\
  &&\hspace{-0.29cm}
  \Big\langle \, {\rm P} \, {\rm{exp}} 
  [ -\frac{ig}{2} \int_{\it{P_+}}
  d\sigma^{\mu \nu}
  \, \hat{G}_{\mu \nu}^{a} 
  (\frac{\lambda^{a}}{2} \otimes 1) ]
  \nonumber\\
  &&\hspace{-0.0cm}
  {\rm P} \, {\rm{exp}} [ -\frac{ig}{2} \int_{\it{P_-}}
  d\sigma^{\mu \nu}
  \, \hat{G}_{\mu \nu}^{a} (1 \otimes \frac{\lambda^{a}}{2}) ]
  \; \Big\rangle_G \Big\}.
  \label{idea}
\end{eqnarray}
%
%
%
\noindent
Introducing a total shifted
field strength tensor $\hat{G}_t$ as
\begin{equation}
  \hat{G}_{t,\mu \nu}(o,x;C_x)
  = \left\{
    \begin{array}{l@{\quad}l}
      \hat{G}^a_{\mu \nu}(o,x;C_x)
      (\frac{\lambda^a}{2} \otimes  1) \\
      {\rm{for}}\quad x \,\epsilon \,
      P_+\\
      \hat{G}^a_{\mu \nu}(o,x;C_x)
      (1 \otimes \frac{\lambda^a}{2} ) \\
      {\rm{for}}\quad x \, \epsilon \, 
      P_-
    \end{array}
  \right.
  \label{gtot}
\end{equation}
%
%
we can rewrite the two exponentials in (\ref{idea})
as one exponential defined in the direct product space.
In this way we get from (\ref{stokes}) a path-ordered integral
over the double  pyramid mantle $P=P_++P_-$:
\bear
  &&\hspace{-0.69cm}
    \Big \langle W_+^M \, W_-^M \Big \rangle_G =
  \nonumber\\
  &&\hspace{-0.29cm}
  \frac{1}{9} {\rm{Tr}}_2 \Big\langle 
  {\rm P}\ {\rm{exp}} [ -i\frac{g}{2}
  \int_{P}
  d\sigma (x)  \, \hat{G}_t (x) ]
  \Big\rangle_G .
  \label{collected}
\ear
%
%
Here and in the following
we suppress the Lorentz indices if there is no confusion. Note that
the path orderings on $P_+$ and $P_-$ do not interfere
with each other. Thus the path-ordering on $P$ can for instance
be chosen such that all points of $P_+$ are ``later''
than all points of $P_-$.

For the expectation value of the
single surface ordered exponential (\ref{collected}) we 
make a matrix cumulant expansion as explained 
in (2.41) of \cite{na96}:
\begin{eqnarray}
  &&\hspace{-0.69cm}
  \Big\langle {\rm P}\, {\rm{exp}} [ -i\frac{g}{2}
  \int_{P}
  d \sigma (x)  \, \hat{G}_t (x) ] \Big\rangle_G = 
  \nonumber\\
  &&\hspace{-0.39cm}
  {\rm{exp}} \Big[\sum_{n=1}^{\infty} 
  \frac{1}{n !}
  (-i\frac{g}{2})^n
  \int d \sigma (x_1) \cdots d\sigma (x_n) 
  \cdot
  \nonumber\\
  &&\hspace{0.3cm}
  K_n(x_1,..,x_n) \;
  \Big] .
  \label{cumulant}
\end{eqnarray}
%
%
Here the cumulants $K_n$ are functional integrals
over products of the
non-commuting matrices $\hat{G}_t$ of (\ref{gtot}). Thus
one has to be
careful with their ordering.

Neglecting cumulants higher than n=2 and 
using in addition
$\langle \hat{G}_{\mu \nu}^a\rangle_G=0$ leads to
\begin{eqnarray}
  &&\hspace{-0.69cm}
  \Big\langle W_+^M \,  W_-^M \Big\rangle_G =
  \frac{1}{9} {\rm{Tr}}_2 \, {\rm{exp}}(C_2(\vec x_T,\vec y_T,\vec b_T)),
  \nonumber\\
  &&\hspace{-0.69cm} 
  C_2 (\vec x_T,\vec y_T,\vec b_T)=
  -\frac{g^2}{8} \int_P d\sigma(x_1)
  \int_P d\sigma(x_2) \cdot
  \nonumber\\
  &&\hspace{0.9cm}
  \Big\langle {\rm P} (\hat{G}_t(o,x_1;C_{x_1})
  \hat{G}_t(o,x_2;C_{x_2}))
  \Big\rangle_G 
  \label{mcum}
\end{eqnarray}
%
%
where $C_2$ is a $9\times 9$ matrix, invariant under SU(3)
colour rotations.

Now we use the MSV ansatz \cite{msv}
for the correlation function of two
shifted field strengths $\hat{G}_{\mu \nu}^a$ which consists of two
Lorentz tensor structures multiplied by invariant functions $D$ times
$\kappa$ ($0$ $\leq$ $\kappa$ $\leq 1$)
and $D_1$ times (1--$\kappa$) respectively.
In QCD lattice measurements show \cite{lat1} $\kappa$$\sim$3/4
whereas in an abelian theory $\kappa$$=$$0$. For deriving
confinement in terms of the MSV $\kappa$$\neq$$0$ is crucial
\cite{msv}.
We find that $\kappa$$\neq$$0$ is also necessary to
reproduce the experimental data for $d\sigma/dt$ \cite{naber}.
Further parameters of the MSV are 
the gluon condensate $G_2$ and the
vacuum correlation length $a$.

Inserting this ansatz in $C_2$ we get \cite{naber}
\begin{eqnarray}
  &&\hspace{-0.69cm}
  \Big\langle W_+^M(\frac{1}{2} \vec{b}_T,\vec{x}_T) \,
  W_-^M (-\frac{1}{2} \vec{b}_T,\vec{y}_T) \Big\rangle_G =
  \nonumber\\
  &&\hspace{-0.69cm}
  \; \; \; \frac{1}{9}\; {\rm{Tr}}_2 \; {\rm{exp}}
  [ (\frac{\lambda^a}{2} \otimes \frac{\lambda^a}{2}) \,
  (-i) \, \chi(\vec{x}_T,\vec{y}_T,\vec{b}_T)\;]=
  \nonumber\\
  &&\hspace{-0.69cm} \; \; \;
  \frac{1}{3} e^{i\frac{2}{3}\chi} + 
  \frac{2}{3} e^{-i\frac{1}{3}\chi}.
  \label{cummatrix}
\end{eqnarray}
%
%
%
where $\chi$ is a real function. 
In the last step  one introduces  
projectors $P_s$, $P_a$
satisfying
($\frac{\lambda^a}{2}$$\otimes$$\frac{\lambda^a}{2}$)$=$$\frac{1}{3} 
P_s$--$\frac{2}{3} P_a$.

\section{Meson-meson amplitude}

Inserting (\ref{cummatrix}) in (\ref{ampl}) and using the
mesonic overlap functions 
$w^M_{i,j}(\vec{z}_T)$=$1/(2\pi S_H^2)
exp(-\vec{z}_T^2/2S_H^2)$ 
of \cite{dfk}
our final result for the meson-meson scattering amplitude 
reads
\begin{eqnarray}
  &&\hspace{-0.69cm}
  T_{fi} = (2is) \, (2\, \pi)
  \int_0^{\infty} db \, b \, J_0(\sqrt{|t|}\, b)
  \hat{J}_{M,M}(b),
  \nonumber\\
  &&\hspace{-0.69cm} 
  \hat{J}_{M,M}(b)=
  -\int d^2x_T  \int d^2y_T \,
  w_{3,1}^M(\vec{x}_T) \, w_{4,2}^M(\vec{y}_T)
  \nonumber\\
  && \hspace{1.05cm}\cdot
  \Big\{ \;  
  \frac{2}{3} \cos (\frac{1}{3}  \chi\,) +
  \frac{1}{3} \cos (\frac{2}{3}  \chi\,) - 1
  \,  \Big\} .
  \label{mres}
\end{eqnarray}
%
%
%
Here $J_0$ is the zeroth-order Bessel function. The
sine terms which one would expect from (\ref{cummatrix})
are averaged out by integrating over $\vec{x}_T,\, \vec{y}_T$
because we have for example 
$\chi (-\vec{x}_T, \vec{y}_T, \vec{b}_T)=
 -\chi (\vec{x}_T, \vec{y}_T, \vec{b}_T) $. 

As a consequence
(\ref{mres}) is invariant under the replacement of one hadron by its
antihadron: The exchange of all partons by its antipartons
for a given loop configuration turns around the loop direction
(Fig. 1) which results in  
$\vec{x}_T \rightarrow -\vec{x}_T$. But this leaves the amplitude
(\ref{mres}) invariant.
 
In our approximations, we get only charge conjugation $C=+1$
(pomeron) exchange and no $C=-1$ (odderon) exchange contributions
to the amplitude.
A real part of the amplitude and $C=-1$ exchange contributions
could arise from higher cumulants in
(\ref{cumulant}).

\section{Proton-proton scattering}

Now we come to our results for  $pp$ scattering
where protons are treated in the quark-diquark picture.
First we have to fix the parameters in (\ref{mres}) which
are: 
the QCD vacuum parameters $G_2, \ \kappa$ and $a$ and the proton
extension parameter $S_{H_1}$=$S_{H_2}$=$S_p$. The vacuum
parameters are surely energy and process independent. 
Following \cite{dfk}
the extension parameter $S_p$ will be allowed to
vary with energy. We fix these parameters using as input
experimental data at $\sqrt{s}=23$GeV
for $d\sigma/dt$  and impose in addition
the constraint that our amplitude reproduces the pomeron part
of $\sigma_{tot}$ in the Donnachie-Landshoff (DL) parametrisation
\cite{dola}. We find \cite{naber}: $G_2=(529MeV)^4$, $a=0.32$fm and
$\kappa=0.74$, compatible with lattice determinations \cite{lat2} 
and
$S_p(23{\rm GeV})$=0.87fm which is in the range 
of the
electromagnetic proton radius as it should.
All values compare well with the values determined in previous
work on high energy scattering \cite{dfk,mdoc}. 

$\sigma_{tot}$ increases with
increasing extension parameter.
So in order to calculate $d\sigma/dt$ at higher c.m. energies we fix
$S_p(\sqrt{s})$  requiring again that our model reproduces
the pomeron part of $\sigma_{tot}$ in the DL parametrisation.

Now everything is fixed and we can calculate $\sigma_{tot}$ and
$d\sigma/dt$ from (\ref{mres}). In Fig. 2 we show our results.
%
%
%
\begin{figure}[!t]
  \unitlength1.0cm
  \begin{center}
    \begin{picture}(15.,8.8)

      \put(-0.2,-0.4){

        \epsfysize=9.4cm
        \epsfxsize=8.2cm

        \epsffile{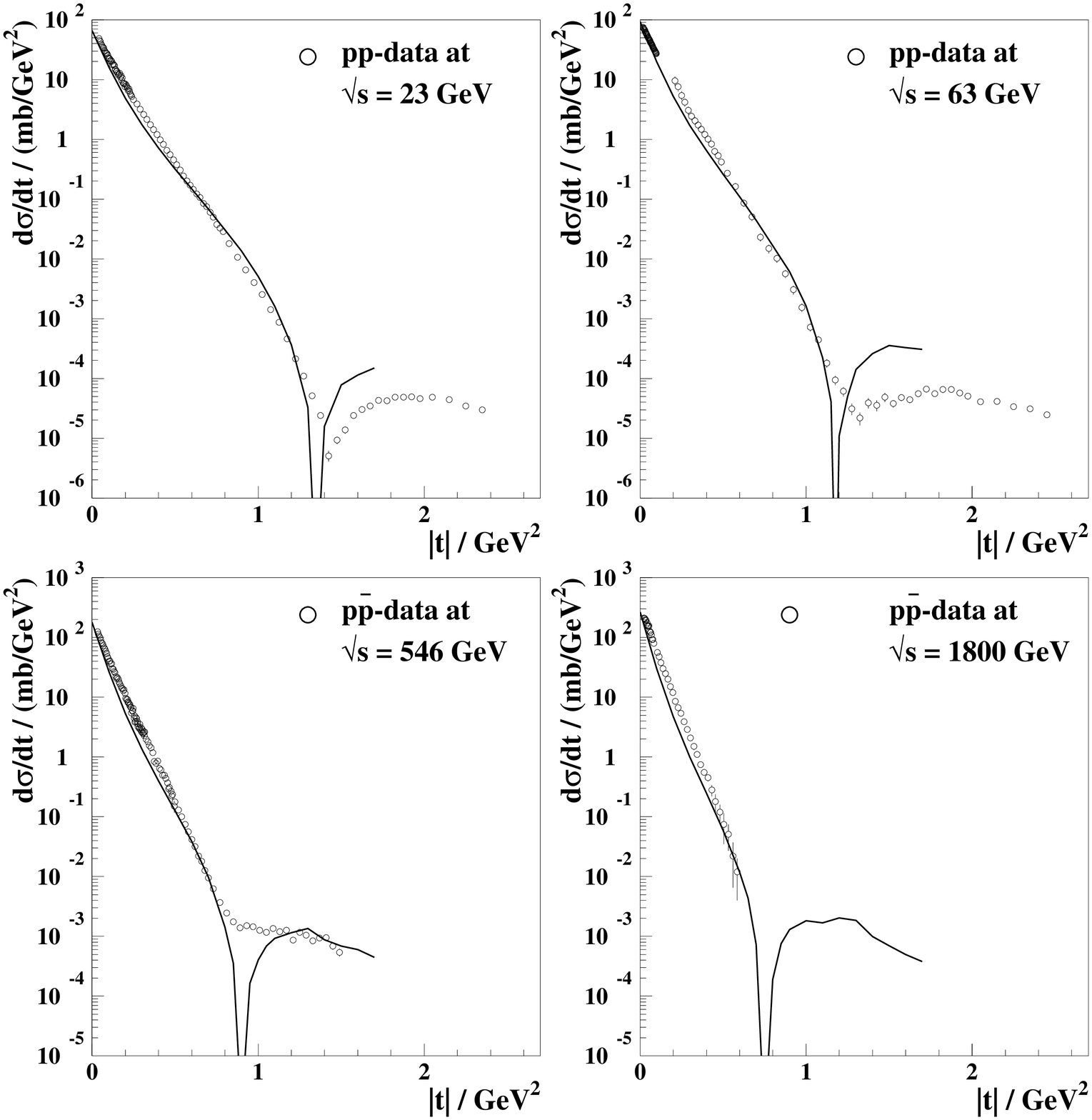}}

    \end{picture}
  \end{center}
\vspace*{-1.1cm}
\caption[a]{ 
Differential elastic cross sections for
c.m. energies $\sqrt{s}=23,63,546$ and $1800${\rm{ GeV}}.
The references of the experimental data are given in 
\cite{naber}.}
  \label{probild}
\end{figure}
%
%
From there one can see
that for all energies
the calculated differential
distributions follow the experimental data quite well over
many orders of magnitude. The fact that this is true  up
to  $\sqrt{s} = 1800 $ GeV supports
the description of the s-dependence by a
$s$-dependent extension parameter $S_p(\sqrt{s})$.
For all energies the imaginary part of our amplitude changes sign
at some $t<0$. Due to the absence of a real part in
(\ref{mres}) the calculated
differential cross sections have a zero there.
This causes an infinitely deep dip in our $t$-distributions.
We expect this dip to be at least partly filled up
once we change to more general quark
configurations and include higher cumulant terms.
The point at which the zero occurs in our calculation
moves to smaller values of
$|t|$ with increasing energy and is always in the region where
experiments see a marked structure.

Finally we want to stress again, that our results for $d\sigma/dt$
depend crucially on
the $\kappa$-term of the MSV,
which  implies in low
energy phenomenology a string tension
$\rho$$\neq$$0$ and so confinement \cite{msv}.
A detailed discussion of this point can be 
found in \cite{naber}.

Of course our model is not perfect.
Our amplitude is purely imaginary and
thus does not satisfy the relation between the phase and
the $s$-dependence required by analyticity
and Regge theory \cite{regge}. Also our
$d\sigma/dt$ is the same for $pp$ and $p\bar p$ scattering.
Experiments show that this is not true
in the dip regions.
We will
have to see if higher cumulant terms and/or a departure from
the strict quark-diquark picture of the proton
will lead us to an improvement on these points 
in our model.

{\bf Acknowledgements:} Thanks to S. Narison
for this very pleasant and 
interesting conference.

\end{document}